%% file: ma1011.tex
\documentstyle[epsf,graphics]{mn}
\newif\ifAMStwofonts

\ifoldfss
  \ifCUPmtlplainloaded \else
    \NewTextAlphabet{textbfit} {cmbxti10} {}
    \NewTextAlphabet{textbfss} {cmssbx10} {}
    \NewMathAlphabet{mathbfit} {cmbxti10} {} 
    \NewMathAlphabet{mathbfss} {cmssbx10} {} 
  \fi
  \ifAMStwofonts
    \ifCUPmtlplainloaded \else
      \NewSymbolFont{upmath} {eurm10}
      \NewSymbolFont{AMSa} {msam10}
      \NewMathSymbol{\upi}     {0}{upmath}{19}
      \NewMathSymbol{\umu}     {0}{upmath}{16}
      \NewMathSymbol{\upartial}{0}{upmath}{40}
      \NewMathSymbol{\leqslant}{3}{AMSa}{36}
      \NewMathSymbol{\geqslant}{3}{AMSa}{3E}

    \fi
  \fi
\fi 

\ifnfssone
  \newmathalphabet{\mathit}
  \addtoversion{normal}{\mathit}{cmr}{m}{it}
  \addtoversion{bold}{\mathit}{cmr}{bx}{it}
  \newmathalphabet{\mathbfit} 
  \addtoversion{normal}{\mathbfit}{cmr}{bx}{it}
  \addtoversion{bold}{\mathbfit}{cmr}{bx}{it}
  \newmathalphabet{\mathbfss} 
  \addtoversion{normal}{\mathbfss}{cmss}{bx}{n}
  \addtoversion{bold}{\mathbfss}{cmss}{bx}{n}
  \ifAMStwofonts
    \ifCUPmtlplainloaded \else
      \UseAMStwoboldmath
      \makeatletter
      \new@mathgroup\upmath@group
      \define@mathgroup\mv@normal\upmath@group{eur}{m}{n}
      \define@mathgroup\mv@bold\upmath@group{eur}{b}{n}
      \edef\UPM{\hexnumber\upmath@group}
      \new@mathgroup\amsa@group
      \define@mathgroup\mv@normal\amsa@group{msa}{m}{n}
      \define@mathgroup\mv@bold\amsa@group{msa}{m}{n}
      \edef\AMSa{\hexnumber\amsa@group}
      \makeatother
      \mathchardef\upi="0\UPM19
      \mathchardef\umu="0\UPM16
      \mathchardef\upartial="0\UPM40
      \mathchardef\leqslant="3\AMSa36
      \mathchardef\geqslant="3\AMSa3E
    \fi
  \fi
\fi 

\ifnfsstwo
  \DeclareMathAlphabet{\mathbfit}{OT1}{cmr}{bx}{it}
  \SetMathAlphabet\mathbfit{bold}{OT1}{cmr}{bx}{it}
  \DeclareMathAlphabet{\mathbfss}{OT1}{cmss}{bx}{n}
  \SetMathAlphabet\mathbfss{bold}{OT1}{cmss}{bx}{n}
  \ifAMStwofonts
    \ifCUPmtlplainloaded \else
      \DeclareSymbolFont{UPM}{U}{eur}{m}{n}
      \SetSymbolFont{UPM}{bold}{U}{eur}{b}{n}
      \DeclareSymbolFont{AMSa}{U}{msa}{m}{n}
      \DeclareMathSymbol{\upi}{0}{UPM}{"19}
      \DeclareMathSymbol{\umu}{0}{UPM}{"16}
      \DeclareMathSymbol{\upartial}{0}{UPM}{"40}
      \DeclareMathSymbol{\leqslant}{3}{AMSa}{"36}
      \DeclareMathSymbol{\geqslant}{3}{AMSa}{"3E}
    \fi
  \fi
\fi 

\ifCUPmtlplainloaded \else
  \ifAMStwofonts \else 
    \def\upi{\pi}
    \def\umu{\mu}
    \def\upartial{\partial}
  \fi
\fi

\title[The massive double-lined O-type binary HD 93205]{Optical spectroscopy of X-Mega targets in the Carina Nebula.\\
       II. The massive double-lined O-type binary HD~93205}

\author[Morrell et al.]
{
N.I. Morrell$^1$\thanks{Member of Carrera del Investigador Cient\'{\i}fico, CONICET, Argentina},
R.H. Barb\'a$^1${\Huge$^{\star}$},
V.S. Niemela$^1$\thanks{Member of Carrera del Investigador Cient\'{\i}fico, CIC, Prov. de Buenos Aires, Argentina},
M.A. Corti$^1$\thanks{Fellow of FOMEC},
J.F. Albacete Colombo$^1$,
\newauthor
G. Rauw$^2$\thanks{Postdoctoral Researcher FNRS (Belgium)},
M. Corcoran$^{3,4}$,
T. Morel$^5$,
J.-F. Bertrand$^{6,7}$,
A.F.J. Moffat$^{6,7}$,
N. St-Louis$^{6,7}$
\\
$^1$ Facultad de Ciencias Astron\'omicas y Geof\'{\i}sicas, Universidad Nacional de La Plata, Paseo del Bosque S/N, B1900FWA La Plata, Argentina\\
$^2$ Institut d'Astrophysique et de G\'eophysique, Universit\'e de Li\`ege,
5, Avenue de Cointe, B 4000 Li\`ege, Belgium\\
$^3$ Universities Space Research Association, 7501 Forbes
Blvd, Ste 206, Seabrook, MD 20706, USA\\
$^4$ Laboratory for High Energy Astrophysics, Goddard Space Flight Center, 
  Greenbelt MD 20771, USA\\
$^5$ Inter-University Center for Astronomy and Astrophysics (IUCAA), 
Post Bag 4, Ganeshkhind, Pune, 411 007, India.\\
$^6$ D\'epartement de physique,
Universit\'e de Montr\'eal, C.P. 6128, Succursale Centre-ville,
Montr\'eal, QC, H3C 3J7, Canada\\
$^7$ Observatoire du Mont M\'egantic, Canada \\
}

\date{Accepted 2000 December 15.
      Received 2000 December 14;
      in original form 2000 October 11}

\pubyear{2000}

\begin{document} 

\maketitle

\begin{abstract}
A new high-quality set of orbital parameters for the O-type 
spectroscopic binary HD~93205 has been obtained combining \'echelle and 
coud\'e CCD observations. The radial velocity orbits derived from the 
He\,{\sc ii} $\lambda$4686\,\AA\ (primary component) and He\,{\sc i} 
$\lambda$4471\,\AA\ (secondary component) absorption lines yield 
semiamplitudes of $133\pm2$ and $314\pm2$ km\,s$^{-1}$ for each binary 
component, resulting in minimum masses of 31 and 13 M$_\odot$ 
($q = 0.42$). 
We also confirm for the binary components the spectral classification 
of O3\,V~+~O8\,V previously assigned. Assuming for the O8\,V 
component a ``normal'' mass of 22 -- 25 M$_\odot$ we would derive for the 
primary O3\,V a mass of ``only'' 52 -- 60 M$_\odot$ and an inclination 
of about $55^\circ$ for the orbital plane.  
We have also determined for the first time a period of apsidal motion 
for this system, namely 185 $\pm$ 16 years using all available radial 
velocity data-sets of HD~93205 (from 1975 to 1999).
Phase-locked variations of the X-ray emission of HD~93205 consisting of
a rise of the observed X-ray flux near periastron passage, 
are also discussed.
\end{abstract}

\begin{keywords}
stars: binaries -- stars: early-type -- stars: individual: HD~93205 --
 X-rays: stars
\end{keywords}

\section{Introduction}
The Carina Nebula region is known to harbor some of the youngest and most 
massive O-type stars in our galaxy (cf. Walborn 1995). 
This region, containing the open clusters Trumpler 14, Trumpler 16 and
Collinder 228, is indeed  the major concentration of O-type stars known 
in the nearby Milky Way. 
Hot stars are frequently observed as X-ray sources, though the mechanism
responsible for this emission still remains not well understood. 
While X-ray emission can arise from the winds of hot single stars, the
binary nature of some of them might also contribute to  X-ray production 
through wind-wind collision effects.
In general, the ratio between X and bolometric luminosities 
($L_{\rm x} / L_{\rm bol}$) seems to be close to $10^{-7}$ for OB stars 
(Corcoran 1999) but the proposed relation exhibits a large scatter. 
With the aim of addressing some of those problems, in 
late 1995, the {\em X-Mega}
collaboration  (Corcoran et al., 1999) started observations and analysis of 
X-ray emission from hot massive stars. Obviously, the Carina Nebula was 
one of its first selected targets.

In the first paper of this series (Albacete Colombo et al. 2000) we reported
the discovery of a new double-lined O-type binary among the members
of the open cluster Trumpler~16.
Another O-type member of Trumpler 16, HD~93205, is the only O3\,V star 
in the Milky Way known to belong to a double-lined binary system 
(Walborn 1971, 1973). Although similar systems have been recently 
discovered in the LMC (e.g. Massey \& Hunter 1998; Bertrand, St-Louis \&
Moffat 1998), 
HD~93205 is still the earliest type star in our Galaxy for which an orbital 
solution is available. This makes this system specially valuable for
the mass-luminosity relation for the most massive stars.

\input{instr.tex}

Conti \& Walborn (1976, hereafter CW76) classified the binary components of 
HD~93205 as 
O3\,V+O8\,V and presented the first radial velocity orbit for this system, 
deriving a period of $6.0810\pm0.0007$ days and minimum masses of 39 and 15 
M$_{\odot}$ respectively. They found a highly eccentric orbit ($e=0.49$)
which also suggests that this system is extremely young, with both components 
very close to the ZAMS.
Subsequent radial velocity studies by Levato et al. (1991) and Stickland \& 
Lloyd (1993, hereafter SL93) essentially confirmed the preliminary results by 
CW76.

The first observational efforts
aimed to the search of photometric variability in HD~93205
 (van Genderen et al. 1985a, b, 1989) 
 showed only marginal evidence for light variations. 
 Then, Antokhina et al. (2000) found clear phase-dependent
variations with full amplitude $\sim$ 0.02 mag. in visual light.
These are probably related to tidal distortions rather than eclipses, 
leaving room for a wide range of possible orbital inclinations, 
$35^{\circ}  < i < 75^{\circ}$, with  most likely value $i = 60^{\circ}$.
The relatively low values derived for the minimum masses of the binary 
components also support the idea that the inclination is not very  high. 

Several studies considered the mass ratio ($q\sim0.4$) of the HD~93205 binary 
system to be ``anomalous'' (e.g. SL93; Penny et al. 1998). This
statement is based on the assumption that the secondary, an O8\,V star, has a 
mass of $\sim25$ M$_{\odot}$, as suggested by both binary star observations
(cf. Sch\"onberner \& Harmanec, 1995; Burkholder, Massey \& Morrell 1997) 
and evolutionary tracks (cf. Schaller et al. 1992). Then the mass of the O3\,V 
star would be ``only'' $\sim60{\rm M}_{\odot}$, much lower than the 
expected value, based on evolutionary tracks.
 
Taking into account that HD~93205 is, in our Galaxy, the only 
well studied binary system containing an
O3\,V-type non evolved component, and in view of the
apparent discrepancy between the observed mass ratio and that predicted by
the evolutionary models, we decided to perform a new study mainly based on 
high resolution CCD spectroscopic observations of this key binary system.
Moreover, HD~93205 is, as mentioned above, one of the Carina targets observed
 in the context of the {\em X-Mega} campaign, and a new orbital determination
was advisable in order to plan and interpret the X-ray observations.

\section{Observations}
The present study is the result of a collaboration in which several data 
sets were obtained at three different southern observatories, using 4 
telescopes and 5 spectrographs. Table~\ref{instr} lists details of each 
instrumental configuration.

Thirty-nine CCD \'echelle spectra of HD~93205  were obtained at the Complejo 
Astron\'omico El Leoncito\footnote{CASLEO is operated under agreement 
between CONICET and National Universities of La Plata, C\'ordoba and San Juan}, 
Argentina (CASLEO) between 1995 and 1999 with the 2.15-m Jorge Sahade 
Telescope. We used a REOSC \'echelle Cassegrain spectrograph and a Tek 
$1024\times1024$ pixel CCD as detector to obtain thirty-five 
spectra in the approximate  wavelength range 3500 to 6000\,\AA, at a 
reciprocal dispersion of 0.17\,\AA\,px$^{-1}$ at 4500\,\AA. 
Four \'echelle spectra of HD~93205 were obtained at CASLEO in January~1995 
with the configuration described above but binning the CCD by a factor 2.
Four additional observations were obtained at CASLEO with a Boller \& Chivens 
(B\&C) spectrograph attached to the 2.15-m telescope, using a PM $512\times512$ 
CCD as detector, and a 600\,l\,mm$^{-1}$ diffraction grating, yielding a 
reciprocal dispersion of 2.5\,\AA\,px$^{-1}$.

Sixteen CCD spectra of HD~93205 were obtained at the European Southern 
Observatory (ESO) in Chile. Six of them were gathered with the B\&C 
spectrograph attached to the 1.5-m telescope, at a reciprocal dispersion 
of 0.6\,\AA\,px$^{-1}$ covering the wavelength region from 3850 to 4800\,\AA.
The detector was a thinned, UV flooded CCD (ESO ccd\#39) and the spectral
resolution as measured from the FWHM of the lines of the helium-argon
calibration spectra was 1.1\,\AA. 
 Ten spectra were obtained with the 1.4-m Coud\'e Auxiliary Telescope (CAT),
    using the Coud\'e Echelle Spectrometer equipped with the Long Camera
    (LC) in 1997 and the Very Long Camera (VLC) in 1998. The detector 
    used during both runs was a Loral $2688 \times 512$ pixel CCD with 
    a pixel size of 15\,$\mu$m $\times$ 15\,$\mu$m. The effective 
    resolving power as derived from the FWHM of the lines of the ThAr
    calibration exposures was 70000. Typical exposure times were of 
    the order of 30-45 minutes. The wavelength domain was 
    centered at $\lambda$\,4470\,\AA, and covered a narrow spectral 
    window of 40\,\AA\ for the LC and 20\,\AA\ for the VLC.   

Nine CCD spectra were obtained with the Cassegrain spectrograph
attached to the  0.6-m telescope at the University of Toronto
Southern Observatory (UTSO), Chile. These data, 
obtained with a 600 l\,mm$^{-1}$ grating blazed at 4700 \AA\, in fist order, 
 cover the blue region of the spectrum at a reciprocal dispersion
 of 2\,\AA\,px$^{-1}$. 
Consecutive exposures  were combined in order to obtain
one radial velocity determination for each observing night, as
shown in Table~\ref{vr_low}.

The usual sets of bias and flat-fields were also secured for each
 observing night. 

Data from CASLEO and UTSO were reduced and analyzed with IRAF\footnote{IRAF 
is distributed by NOAO, operated by AURA, Inc., under agreement with NSF.} 
routines, while MIDAS\footnote{
MIDAS is developed and maintained by the European Southern Observatory} 
software was used for the reductions and analysis of data obtained at ESO.

\section{Radial Velocities}

One of our goals being the determination of an accurate radial velocity orbit 
for the HD~93205 binary system, we decided to use only the high resolution CCD
observations for that purpose. However, all the available spectroscopic
observations, either newly obtained or previously published, were used in
the analysis, leading to a determination of the period of apsidal motion.

Table~\ref{vr_low} shows the journal of our low resolution 
observations. 
These radial velocities were obtained through simple Gaussian fitting to the
observed He\,{\sc i} and He\,{\sc ii} absorption line profiles, considering 
that the major contributor to the former is the O\,8 secondary star, while 
He\,{\sc ii} lines mainly originate in the primary O\,3 component. 

\input{vr_low.tex}

\input{vr_high.tex}

\begin{figure}
\epsfxsize=85mm \epsfbox[90 240 450 540]{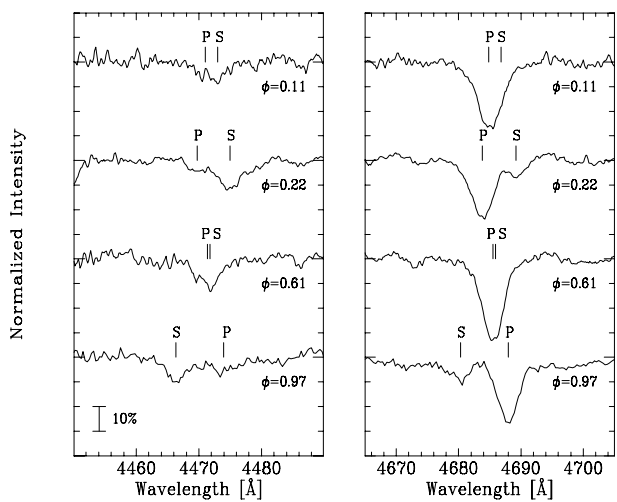}
\caption{Rectified spectrograms of HD~93205 in the region of 
the He\,{\sc i} 4471\AA\ (left) and He\,{\sc ii} 4686\AA\ (right) 
absorption lines at four orbital phases. 
``P'' and ``S'' tick marks show the expected
radial velocities for both binary components.}
\label{he_spec}
\end{figure}

Table~\ref{vr_high} presents the journal of our high resolution observations.
Again, orbital phases were computed with the new ephemeris.
We list heliocentric radial velocities measured for He\,{\sc i} 
$\lambda4471.479$\,\AA\ 
and He\,{\sc ii} $\lambda4685.682$\,\AA\ for both binary components. 
Depending on the degree of blending, we applied simple or simultaneous
 double Gaussian fitting to the absorption line profiles, in order to 
obtain the corresponding radial velocities. 
We selected these two lines for the radial velocity calculations, because they
lie in a region of the CASLEO \'echelle spectra where highest S/N occurs.
The He\,{\sc ii} line was better observed in the primary than in
the secondary component, while the reversal is true for the He\,{\sc i} line.
In addition, He\,{\sc i} 4471\,\AA\ is the only spectral line present
in the CAT data set. 
For those  cases where no double lines were observed, we attributed
the He\,{\sc ii} line to the O3\,V primary and the He\,{\sc i} line
to the O8\,V secondary components, respectively.
Figure~\ref{he_spec} displays the regions containing He\,{\sc ii} 4686\,\AA\ 
and He\,{\sc i} 4471\,\AA\ absorption lines in 
several CASLEO high-resolution \'echelle 
spectrograms at four different phases of the binary period, illustrating the 
contribution of both binary components.

The stability of the radial velocity system for the CASLEO \'echelle
 data was checked in two ways:
a) by radial velocity measurements of nebular emission lines ([O\,{\sc iii}], 
[O\,{\sc ii}], and H\,{\sc i}) present in the spectrum of HD~93205; 
and b) by measurement 
of several spectra of the constant radial velocity star HR~2806 for 
which we obtained a mean heliocentric radial velocity of $26\pm1$\,km\,s$^{-1}$ 
(s.d.) in excellent agreement with other determinations (28.0\,km\,s$^{-1}$, 
Garmany, Conti, \& Massey 1980; 25.9\,km\,s$^{-1}$, Penny et al. 1993; 
27.2\,km\,s$^{-1}$, Garc\'{\i}a et al. 1998). A typical internal standard deviation 
for our measurements of the radial velocity of HR~2806 is 4\,km\,s$^{-1}$.

\section {Orbital Elements and their Discussion}
\subsection {The radial velocity orbit}

The first determination of the orbital period of HD~93205 was performed
by CW76, with a value of $P = 6.0810 \pm 0.0007$ days.
Subsequent determinations by Levato et al. (1991) and SL93 gave values of
 $P = 6.08071 \pm 0.00007$ and $P = 6.08081 \pm 0.00007$ 
days, respectively,  using all the radial velocity data available for the 
primary star at that moment. However the best period found by SL93 to fit 
the {\em IUE} data is $P=6.0821 \pm 0.0004$ days. These authors 
also mention a possible apsidal motion with a period of about 400 years, but 
the data then available were not enough for a conclusive result.

\begin{figure}
\epsfysize=65mm\rotatebox{0}{\epsfbox{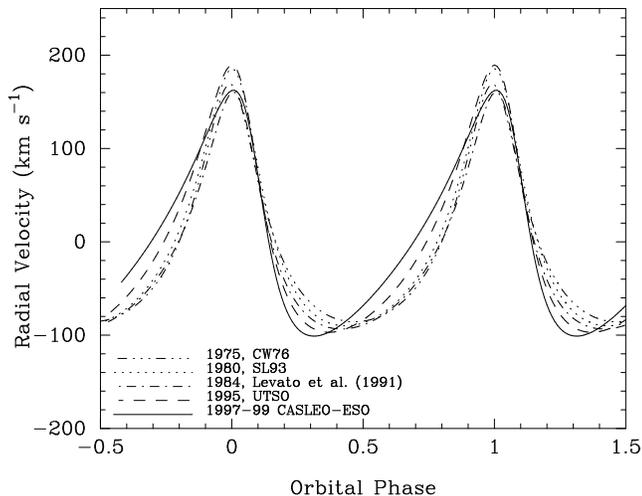}}
\caption{Best fits to radial velocities of the primary star in HD~93205 at
different epochs to display the change of shape of the radial velocity
curve. Solutions were derived adopting a $P=6\fd0803$ and a fixed eccentricity 
$e=0.37$. Different radial velocity curves are shifted in phase to put all
radial velocity maxima in phase $\phi=0\fp0$.
All curves have been shifted to the same systemic velocity.
Sources are identified according to  the epoch of observations 
and authors.}
\label{vrvar.ps}
\end{figure}

Apsidal motion changes the shape of a binary's radial velocity curve 
(in the elliptical case) when observed at different epochs. 
This effect in the orbit of HD~93205 is illustrated in Figure~\ref{vrvar.ps} 
where each curve represents the best fit to the radial velocities of HD~93205 
obtained during different years since the discovery of its binary nature,
namely, CW76, Levato et al. (1991), SL93, 
data from UTSO (this paper, 1995), CASLEO and ESO data (this paper,
1997 - 1999). We ploted each data set using the ephemeries derived 
from the last set (1997 - 1999).

In principle, apsidal motion also precludes the successful determination of 
the orbital period of a binary with the usual simple period searching
 techniques applied to radial velocity data gathered over several years. 
The fact that the various datasets have quite different qualities 
adds more confusion to this point. After some trials,
which comprised the application of the Lafler \& Kinman (1965)
method and several subsequent modifications of it, e.g.
 Cincotta, M\'endez \& N\'u\~nez (1995), we concluded that we cannot attain, 
for a simple determination with constant orbital parameteres 
an accuracy higher than 0.0004 day for the period.
We therefore adopted for the following calculations 
the most probable period derived from the high resolution datasets, namely 
$6.0803\pm0.0004$ days.

\begin{figure*}
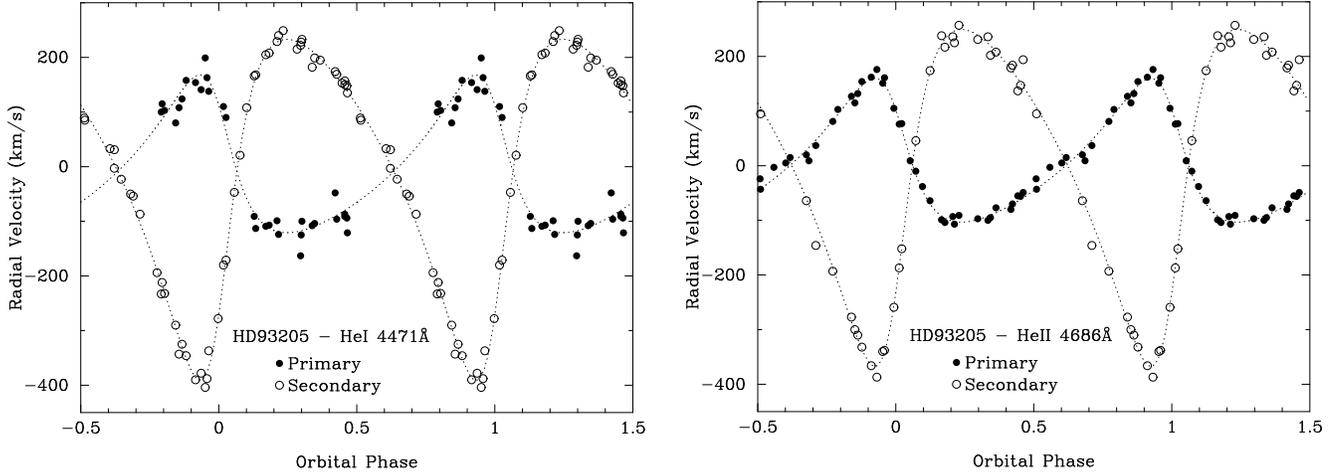

\begin{minipage}{180mm}
\epsfysize=85mm\rotatebox{-90}{\epsfbox{VR_HeI.sol.ps}}
\vskip-62.5mm\hskip90mm\epsfysize=85mm\rotatebox{-90}{\epsfbox{VR_HeII.sol.ps}}
\caption{Left. Observed radial velocity for the primary ($\bullet$) and 
secondary ($\circ$) components of HD~93205 derived from He\,{\sc i} 4471\AA\ 
absorption line. Right.
Observed radial velocity for the primary ($\bullet$) and secondary ($\circ$)
components of HD~93205 derived from He\,{\sc ii} 4686\AA\ absorption line.}
\label{orb_4471_4686}
\end{minipage}
\end{figure*}


We used for the determination of the radial velocity orbit of the HD~93205 
system the He{\sc ii}~4686 and He{\sc i}~4471 lines of both binary components
measured on the high resolution observations listed in Table~\ref{vr_high}.
The orbital elements were obtained using a modified version of the code
originally written by Bertiau \& Grobben (1969),
considering  the radial velocity measurements for each binary component
independently, or both binary components together.

The resulting orbital elements are listed in Table~\ref{orb_sol}.

\begin{figure*}
\epsfysize=130mm\rotatebox{-90}{\epsfbox{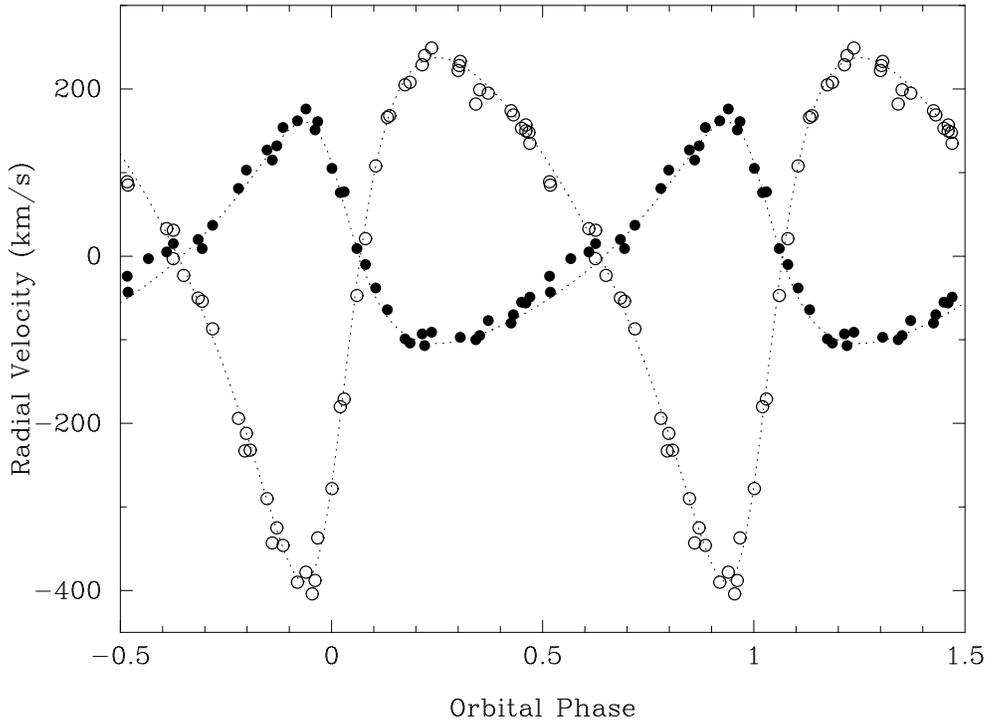}}
\caption{The spectroscopic orbit 
of HD~93205 computed from He\,{\sc ii} 4686\AA\ 
radial velocities for the primary component
 ($\bullet$) and He\,{\sc i} 4471\AA\ 
for the secondary component ($\circ$). This radial velocity curve is labeled as 
{\em combined} solution in Table~\ref{orb_sol}.}
\label{orb_comb}
\end{figure*}

Figure~\ref{orb_4471_4686} shows the observed radial velocities
along with the orbital solutions of Table~\ref{orb_sol} for He\,{\sc i} 4471\AA\
and He\,{\sc ii} 4686\AA\ separately.
Considering that the He\,{\sc ii} lines arise mainly in the O3\,V primary component
and He\,{\sc i} is more representative of the secondary, we preformed an
orbital solution combining the radial velocities of He\,{\sc ii} 4686\AA\
and He\,{\sc i} 4471\AA\  for the primary and secondary components,
 respectively. 
This solution is presented in Table~\ref{orb_sol} in the column labeled as
{\em combined} and in Figure~\ref{orb_comb}. 
Previous published solutions obtained by CW76 and SL93 are also shown
 in the same Table~\ref{orb_sol} for comparison.

\input{orb_sol.tex}

\subsection {The period of apsidal motion}

\begin{figure}
\epsfysize=65mm{\epsfbox{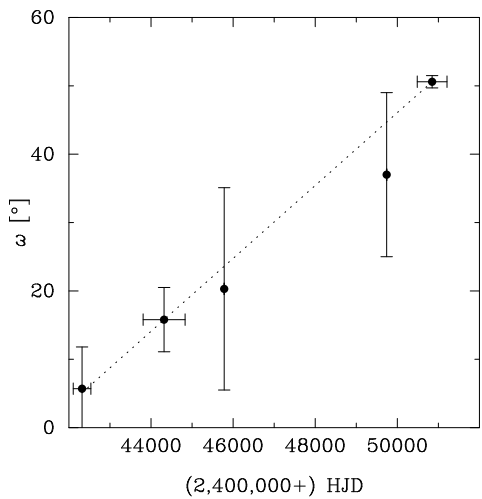}}
\caption{Variation of longitude of periastron ($\omega$) as a function of time. 
Horizontal error bars 
show the time elapsed for each dataset. Vertical error bars correspond to the 
errors obtained for each $\omega$ considering a fixed eccentricity
 ($e=0.37$) derived
in the orbital solution from radial velocities measured in our high-resolution 
spectrograms. The dotted line is the least-square fitting with a slope of 
$0\fdg00533\pm0\fdg00051$ per day ($0\fdg0324\pm0\fdg0031$ per orbital cycle).}
\label{omega.ps}
\end{figure}

An inspection of the new data  suggests a somewhat lower eccentricity and a 
considerably larger value of the longitude of periastron
 $\omega$ than those found in 
previous solutions. While the lower eccentricity probably originates
 in a better phase coverage of our observations, the change in $\omega$
 must be due to apsidal motion (as first suggested by SL93).
Consequently, we decided to investigate the hypothesis of apsidal motion,
 calculating orbital solutions for each individual radial velocity data-set
 obtained over the years 
since the first determination by CW76. We used the radial velocity 
data for the primary component only, and adopted for the orbital period and 
eccentricity our best estimates of $P = 6.0803$~days and $e=0.37$. 
Table~\ref{omega.tab} presents the resulting $\omega$ for the corresponding
 data-sets. 
The variation of $\omega$ as a function of time is represented
 in Figure~\ref{omega.ps} 
along with the obtained least-square linear fit to those data, yielding an
 apsidal motion rate of $\dot{\omega}=0\fdg0324\pm0\fdg0031$ per cycle, which
corresponds to a period of $185\pm16$ years. The ratio of apsidal to orbital
motion ($\sim10\,000$) is of the same order as observed for other
 massive binary systems, e.g. $\iota$ Ori (Marchenko et al. 2000),
 EM~Car (Andersen \& Clausen 1989), AO~Cas (Monet 1980).
Assuming for the HD~93205 binary system a total mass of the order
 of 100~M$_{\odot}$, 
which seems reasonable considering the spectral types of the components,
 the expected relativistic contribution to the apsidal motion would be 
(following Batten, 1973) close to $\dot{\omega}=0\fdg004$ per cycle.
As the observed value is almost 10 times larger, we conclude that the
 apsidal motion of HD~93205 cannot be accounted for by relativistic effects.
 Tidal and rotational distortions must be the main cause of  the apsidal 
motion in this binary.  The possibility that this effect is partially
due to interaction with a third body in the system, cannot be ruled out. 
But if present, such a third body should probably be massive enough to be
 detected spectroscopically, which is not the case.

\input{omega.tex}

\subsection {The mass ratio $q$ and the spectral classifications of
the binary components}

We now consider the mass ratio ($q$). Our new value
confirms those obtained in previous investigations, as seen from the 
coincidence (within the corresponding error bars) of the derived
 radial velocity semiamplitudes for both components. 
As pointed out by Penny et al. (1998), the mass 
ratio of HD~93205 is the most problematic parameter for this binary system
 because it results in a surprisingly low mass of $51-58$\,M$_\odot$ for
 the primary O3 component when a reasonable value (around  $22-25$\,M$_\odot$)
 is adopted for the secondary O8 component (in such a case the orbital 
inclination would be $57^\circ-54^\circ$ thus excluding the possibility
 of detecting photometric variations 
due to eclipses, as also pointed out by Antokhina et al. 2000). 
The suggestion by SL93 that a somewhat
earlier spectral type for the secondary component would help to solve this
problem, is ruled out by Penny et al. (1998) and also by a careful
analysis of our high resolution data. In fact, we inspected each \'echelle
observation in our data-set looking for changes in the spectral
types of the binary components. 
For the secondary component, we confirm the O8\,V spectral classification
assigned since the beginning by  CW76, even though slight 
changes, whose phase-dependence is unclear, might occur
(as can be barely seen in  Figure~\ref{he_spec}). 
Those changes could be related to the Struve-Sahade effect (Howarth et al.
1997) but, since the lines of the secondary component are considerably
 affected by dilution from the higher continuum of the primary component, 
their S/N in our data is not high enough to ensure the detection of 
equivalent width variations.
Consequently, we are persuaded to keep the O8\,V spectral classification 
as most representative of the secondary component of the HD~93205 system.

Regarding the primary O3\,V star, we notice that He\,{\sc i}\,4471\AA\ is 
present in its spectrum, but, as pointed out by Walborn (1999) this fact
does not suffice
to assign a cooler spectral type to the star, because faint He\,{\sc i} lines
are also present on spectra of other O3 stars, for instance, HDE~303308 
observed with the same instrumental set-up, a fact also mentioned by
Walborn \& Fitzpatrick (1990).

\begin{figure}
\epsfysize=55mm{\epsfbox{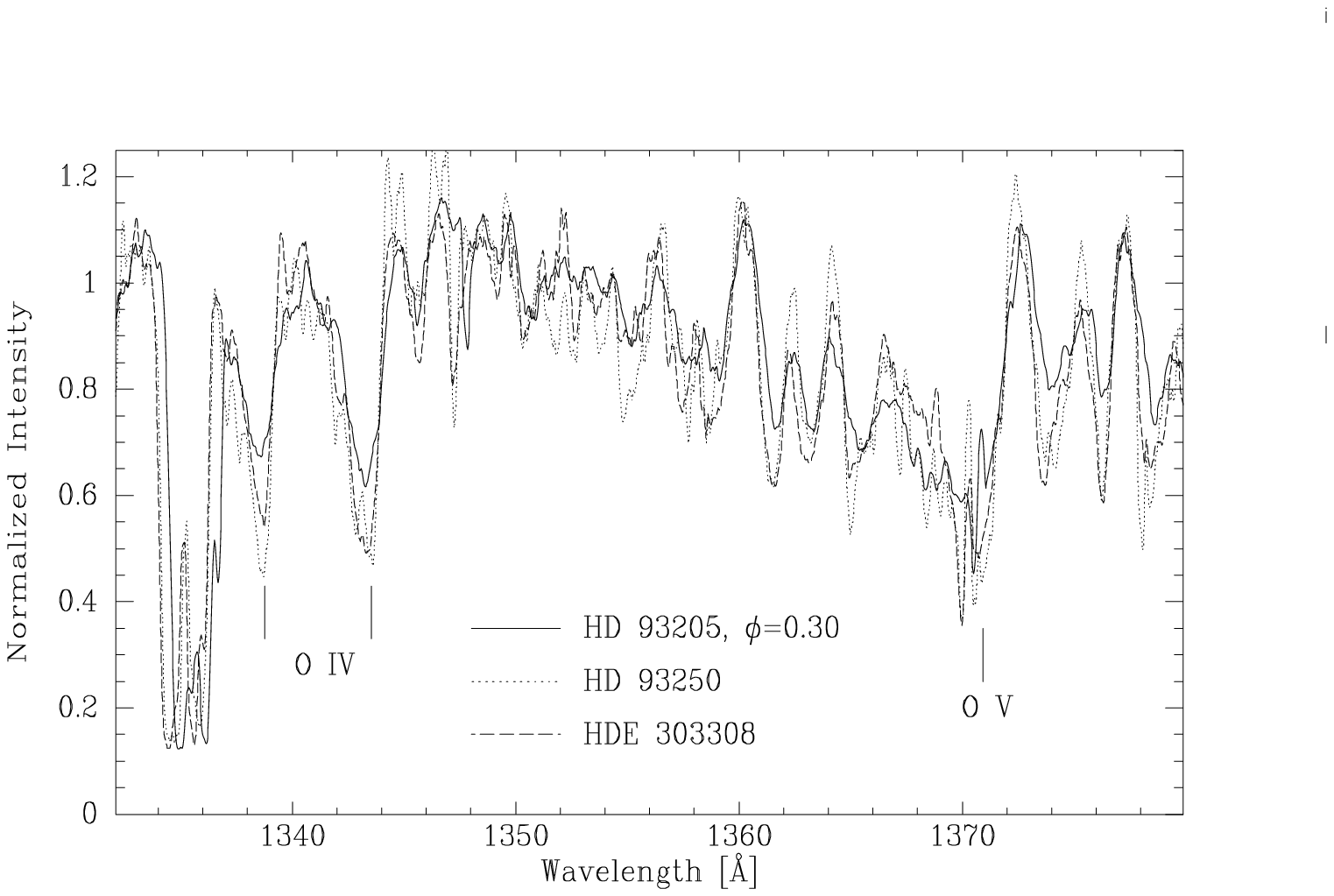}}
\caption{Portion of {\em IUE} spectra of HD~93205 (at orbital
 phase $\phi=0.3$, full
line), HD~93250 (dotted line) and HDE~303308 (dashed line). 
Note that O\,{\sc v} 
1371\AA\ wind profile is very similar in all three stars, 
but O\,{\sc iv} doublet 
is weaker in HD~93205, supporting the O3 classification for the 
primary component.}
\label{iue1}
\end{figure}

We also searched the {\em IUE} database and retrieved all the available
 observations of HD~93205, along with those corresponding to HDE~303308,
 HD~93250, HD~46223 and HD~96715 presented as O3\,V((f)) and O4\,V((f))
 prototypes in the atlas of ultraviolet spectra by Walborn,
Nichols-Bohlin \& Panek (1985).
 A thorough confrontation among those ultraviolet 
spectra showed that the ratio of the stellar wind profile of
 O\,{\sc v}\,1371\AA,
to the  O\,{\sc iv} doublet at $1339-1343$\AA\ is even higher in the spectrum 
of the primary component of HD~93205 than in the spectra of HDE~303308 and
HD~93250, both classified as O3\,V((f)), (see Figure~\ref{iue1}).
Since this ratio is definitely higher for the O3\,V than the O4\,V standard
stars (see Walborn et al. 1985) we conclude that it brings more
evidence in support of the classification of the primary component of
HD~93205 as O3\,V.
 
We also noticed that the feature at 1751\AA\ does not share the binary
motion of HD~93205; thus, it is probably dominated by an interstellar line 
that blends with the N\,{\sc iii} 1751\AA\ absorption line,
used in the above mentioned atlas as a
discriminant between spectral types O3 and O4.
This feature was latter identified as is\,Ni\,{\sc ii} 1751.9\,\AA\
by  Walborn (1999) who also finds that both 1748\,\AA\ and 
1752\,\AA\ N\,{\sc iii} lines are weaker in the O3 spectrum 
than in other O-type stars,
then allowing the interstellar Ni\,{\sc ii} line to dominate.

After reviewing the spectral classification of the binary components of
HD~93205 we confirm, based on all the available material, the
spectral types of O3\,V and O8\,V  assigned earlier, with the
suspicion of slight variations not larger than one sub-class
in the secondary spectrum. This leaves us again with the problem of
explaining a mass ratio incompatible with present stellar evolutionary
models (e.g. Schaller et al. 1992).

\subsection {Position of the binary components in the HR-Diagram}

Once the spectral classification of the binary components is confirmed,
we can use the photometric information in order to place HD~93205
on a theoretical Hertzprung-Russell Diagram (HRD), 
$M_{\rm bol}$ vs $\log T_{\rm eff}$.
We adopt for HD~93205 the distance modulus obtained by Massey \& Johnson
(1993) for Trumpler~16 ($V_0-M_{\rm V}=12.55$). These authors give for HD~93205
an intrinsic visual magnitude $V_0=6.48$ (using $E_{B-V}=0.40$ and
$R=3.2$). In order to derive individual bolometric magnitudes, we need 
an estimate of the visual luminosity ratio of the binary
components. The luminosity ratio was determined through Petrie's 
method as described by Niemela \& Morrison (1988). 
Petrie (1940) demonstrated that the luminosity ratio can be
determined through the ratio of the equivalent widths ($EW$) of well 
separated spectral lines in the binary components, compared to $EW$ 
of single stars of the same spectral types. In order to apply
this method, we measured $EW$ of He\,{\sc ii}\,4686\AA, 4542\AA, 
and He\,{\sc i} 4471\AA\ in several \'echelle  observations at maximum
radial velocity separation (near phases $0\fp0$ and $0\fp3$) for both binary
components of HD~93205, and also we performed similar measurements on 
spectra of HD~93250, (O3\,V((f))) and Tr16-22 (O8\,V) obtained 
for comparison at CASLEO with the 
same instrumental configuration. These {\em EW} have typical
errors ranging from 0.03 to 0.05 \AA. A summary of the measured
{\em EW}s is presented in Table~\ref{ew}, along with {\em EW} determination
 from Mathys (1988) for HD~93250 and HD73882 (O8\,V).
With this information we derived luminosity ratios,  
$L({\rm O8\,V})/L({\rm O3\,V})$ of  $0.47\pm0.22$, $0.24\pm0.06$,
and $0.22\pm0.06$, using He\,{\sc i}\,4471\,\AA, He\,{\sc ii} 4542\,\AA, and 
He\,{\sc ii}\, 4686\,\AA, respectively. The large relative error in the
 luminosity ratio derived from the {\em EW} of He\,{\sc i}\,4471\,\AA\ 
originates in the faintness of this line in the O3\,V spectrum. 
We then decided to adopt for the luminosity ratio of the binary components
 in HD~93205, the average of the ratios obtained from the two
He\,{\sc ii} lines, namely $0.23\pm0.06$. 
This value is lower than $0.4$ determined by Howarth et al. (1997) from cross 
correlation of {\em IUE}  data and also lower than $0.33$ visually estimated
 by CW76 from photographic spectra.

\input{ew.tex}

With the above determined luminosity ratio, we obtained $V_0=6.70$ 
for the primary 
and $V_0=8.30$ for the secondary, resulting in visual absolute magnitudes 
$M_{\rm V}$ of $-5.87$ and $-4.32$ respectively. With $T_{\rm eff}$, 
and bolometric corrections (BC) corresponding to the spectral types 
(Vacca, Garmany \& Shull, 1996), 
we get $M_{\rm bol}$(O3V)$=-10.41$ and $M_{\rm bol}$(O8V)$=-7.87$, and 
$\log T_{\rm eff}$(O3V)=4.71 and $\log T_{\rm eff}$(O8V)=4.58 which places the 
binary components of HD~93205 very near the evolutionary tracks of 
an 85\,M$_\odot$ ZAMS star and a 27 M$_\odot$ star (slightly above the ZAMS)
 respectively, according to Schaller et al. (1992). 
The calibration of $T_{\rm eff}$ and BC of  Schmidt-Kaler (1982)
would place the binary components on the evolutionary tracks corresponding to 
100\,M$_{\odot}$ and 25\,M$_{\odot}$, respectively. 
These results are obviously inconsistent with the observed mass ratio.
 No matter which calibration we choose, the O8\,V component appears near
 the evolutionary track of a 25\,M$_{\odot}$ star, showing quite
good agreement between masses derived through the observation 
of eclipsing binary systems and numerical evolutionary models of single stars,
 while the O3\,V star will always appear above the 85\,M$_{\odot}$ track,
 a value considerably higher than those corresponding to the observed mass 
ratio ($q=0.42$), which would imply $52-60$ M$_{\odot}$.

\begin{figure}
\epsfysize=85mm{\epsfbox{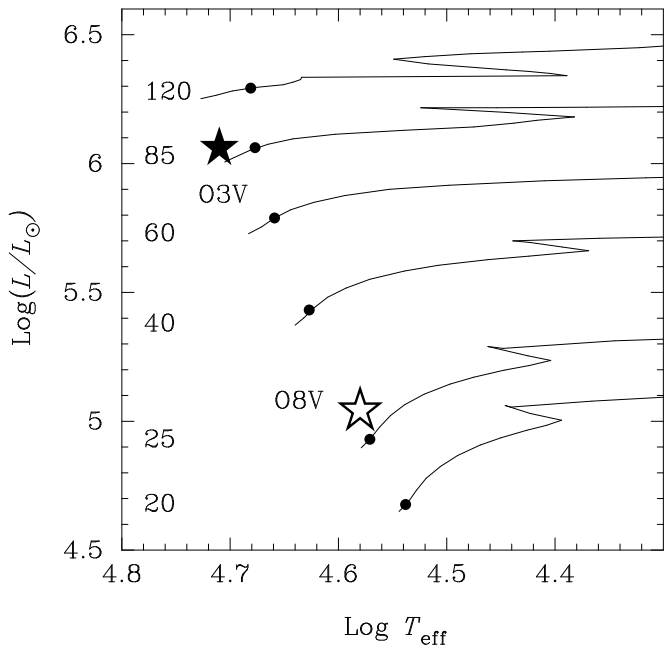}}
\caption{H-R diagram of the binary system HD~93205, 
representing the primary O3\,V 
component with a black star, and the secondary O8\,V with a white star. 
Continuous lines are solar metallicity 
evolutionary tracks from Schaller et al. (1992)and  black dots
a million year isochrone.}
\label{HRD.ps}
\end{figure}

\section{The X-ray Lightcurve of HD 93205}

Since HD 93205 is an eccentric binary with a massive companion, it is
a good candidate to show X-ray emission due to wind-wind collisions
(in addition to intrinsic emission which may be associated with
instabilities in the winds of each star).  In a massive binary, the
wind from the primary star collides with the wind or surface of the
companion, producing hot shocked gas which emits thermal X-rays.
In an adiabatic shock, the X-ray luminosity varies as $1/D$ 
(Stevens, Blondin \& Pollock 1992, Usov 1992), where $D$ is the
separation between the two stars. 
An eccentric binary like HD 93205 should also show phase-locked
emission variations in which the X-ray flux is maximum at periastron
and  minimum at apastron (though in systems in which the colliding
wind shock is eclipsed by the stellar wind, the observed
flux variations are modulated by wind absorption).  For HD 93205,
$e=0.37$, so that the expected ratio of the flux at periastron to the
 flux at apastron is $f_{peri}/f_{ap} \approx (1+e)/(1-e) \approx 2.2$.
The actual variation may be smaller than this due to contamination
by X-ray emission intrinsic to the winds of the component stars.

ROSAT PSPC observations (Corcoran 1995) clearly showed that HD 93205
is a significant X-ray source, though in the PSPC image HD 93205 is
blended with emission from HD 93204, which lies about $20''$ to the
southwest.  A deep ROSAT HRI pointing obtained by the XMEGA group
resolved the emission from HD~93205 and HD~93204, and showed that HD
93204 is about 30\% fainter in the ROSAT band than HD~93205.  In the
resolved HRI image, the average net count rate of HD~93205 is about
0.012 HRI cts s$^{-1}$.
 Assuming a Raymond-Smith type spectrum with
$kT = 1.0$ keV and $N_{H}=3\times10^{21}$ cm$^{-2}$
(typical of other stars in the Carina Nebula), this corresponds
to an observed flux of $3.2\times 10^{-13}$ ergs cm$^{-2}$ s$^{-1}$
and an absorption-corrected flux of about $6 \times 10^{-13}$
ergs cm$^{-2}$ s$^{-1}$ in the broad ROSAT band, 0.2 -- 2.4 keV.
 This corresponds to an unabsorbed luminosity
 $L_{x}=8\times 10^{32}$ ergs s$^{-1}$ assuming $V_{0}-M_{V}=12.55$.
If the total system bolometric luminosity is $L_{bol}\approx 1.3\times 10^6
L_{\odot}$, then $L_{x}/L_{bol}\approx 10^{-7}$, similar to the ratio
for single O stars.  Thus the average luminosity shows no clear evidence of
excess X-ray emission produced by wind-wind collisions.  

Another indication of the importance of colliding wind emission is X-ray
source variability, since single O stars in general do not vary
greatly (Berghoefer et al.  1997), while colliding wind sources
should show phase-dependent X-ray variations.  A ROSAT PSPC lightcurve
of the star was obtained by Corcoran (1996), but this lightcurve
showed little significant variability.  Much more X-ray data is
currently available in the ROSAT archives, so in order to re-examine
whether HD 93205 shows any phase-locked X-ray variability, we
extracted X-ray lightcurves for HD 93205 from archived PSPC and HRI
observations (excluding datasets in which HD 93205 was more than 20$'$
from the center of the ROSAT field, to avoid complications due to
vignetting and due to blending with nearby sources).  We used both
PSPC and HRI pointings, since both instruments are somewhat
complementary: the PSPC provides greater sensitivity, though the
observations suffer more from contamination by HD 93204; the HRI has
poorer sensitivity than the PSPC but does not suffer as much from
source contamination.  Table~\ref{rosat}
  lists the available ROSAT data we used
(those beginning with ``rh'' indicate HRI data, those with ``rp'' PSPC
data), the observer, and the exposure time in seconds.

\input{rosat.tex}
%

\begin{figure}
\epsfysize=45mm{\epsfbox{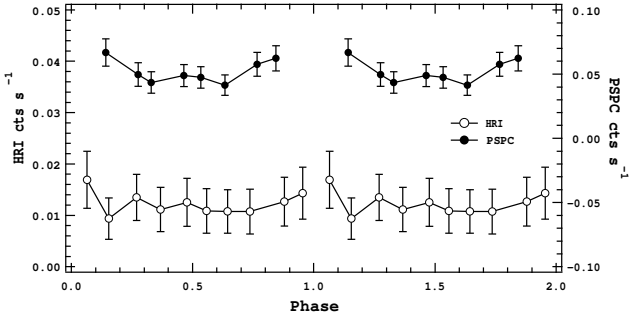}}
\caption {ROSAT PSPC and HRI lightcurves for HD 93205.}
\label{xrays}
\end{figure}

For each dataset we extracted source lightcurves in 512 second bins
from a circular source region of about 45$''$ radius for the PSPC and
about $12''$ radius for the HRI centered on HD 93205.  We extracted
background lightcurves with the same time binning from an apparently
blank circular region centered roughly between the X-ray sources
$\eta$ Carinae, WR 25, HD 93129 and HD 93250, of radius $\approx 1'$.
For each dataset we generated background-subtracted source
lightcurves, and then combined the data for each instrument
individually and phase-averaged the resulting lightcurves using the
``combined'' ephemeris given in Table~\ref{orb_sol}.  The resulting phase-averaged
PSPC and HRI lightcurves are shown in Figure~\ref{xrays}.  Both lightcurves
indicate a rise in X-ray flux from HD 93205 near phase=0, i.e. near
periastron passage.  Though the significance is not high, the
similarity in phasing and amplitude of the HRI and PSPC lightcurves
suggests that at least some of the X-ray emission from HD 93205
originates in the wind-wind collision between the O3 star and the O8
companion.  The ratio of the PSPC rate near periastron ($\phi = 0.9$)
to that at $\phi = 0.5$ is about 1.75.  For the HRI lightcurve, the
ratio of the flux at $\phi=1.08$ to that near $\phi=0.5$ is about 1.70.
Both the PSPC and HRI lightcurves show a level of variability which is
near that expected from a colliding wind model where the X-ray flux
varies as the inverse of the stellar separation.


\section {Summary and Conclusions}

We have presented a new and improved determination of orbital elements for the
HD~93205 binary system based on high resolution CCD spectroscopy.
Our radial velocity orbits yield semiamplitudes of $K_1=133$ km\,s$^{-1}$ 
and $K_2=314$ km\,s$^{-1}$ for the primary and secondary binary 
components, respectively. These values are in the range of the previous
determinations (within the error bar intervals), and therefore they
confirm the mass-ratio $q$ of $0.42$. We have revised the spectral 
classification of both binary components using our optical spectra
and archive UV {\em IUE} observations concluding that the 
O3\,V\,+\,O8\,V spectral types assigned by CW76 
are appropriate, but not excluding the possibility of a slight 
``Struve-Sahade effect'' (as suggested by Howarth et al. 1997).
The luminosity ratio of the binary components, estimated from our
CCD spectra through Petrie's (1940) method, results in
$L({\rm O8V})/L({\rm O3V)}=0.23$. Combining this luminosity ratio with
the published photometry and $\log T_{\rm eff}$ and B.C. derived from
the spectral types, we place the binary components of HD~93205 
on a theoretical  HRD, finding that the O8\,V component lies on the
evolutionary track of a $25-27$\,M$_\odot$ star, and the O3\,V 
component lies between the tracks corresponding to $85-100$\,M$_\odot$.
This would result in a mass-ratio $q=0.25-0.32$, in obvious disagreement 
with  $q$ determined from the radial velocity orbit. 

From studies of  massive binary systems it is well established that the
mass of an O8\,V star is close to $22-25$\,M$_\odot$ (e.g. EM Car, Solivella 
\& Niemela 1986, Andersen \& Clausen, 1989; HH Car, Mandrini et al. 1985,
Sch\"onberner \& Harmanec 1995). Assuming this ``normal'' value of $\sim22-25$ 
M$_\odot$ for the mass of the secondary component of HD~93205, we obtain,
 through 
the determined mass ratio, $52-60$ M$_\odot$ for the O3\,V primary component,
considerably below the mass value predicted by the evolutionary models.
Since HD~93205 is the only double-lined binary system containing an O3\,V
component for which a radial velocity orbit is available at present,
it is not possible to  make any comparisons, but the lack 
of other mass determinations exceeding 60 M$_\odot$ and  our
estimate for the mass of the O3\,V component just obtained, seem to
be significative clues indicating that the derivation of masses
for hot stars based on the comparison of observed luminosities with 
numerical evolutionary tracks might lead to a significant overestimate. 
This seems to support the so-called ``mass
discrepancy'' between masses derived via evolutionary models and
those inferred from observations (mainly spectroscopic), first
studied by Herrero et al. (1992). In a recent  paper,
 Herrero, Puls \& Villamariz (2000) showed that this discrepancy 
(while solved for many cases with the use of new model atmospheres and
evolutinary tracks) still holds for  objects with low gravities,
which is probably not the case of HD~93205.
In another recent paper,  Meynet \& Maeder (2000)
have analysed the effect of rotation on stellar evolutionay models.
These authors state that most part of the mass discrepancies for massive
stars might be solved through the use of rotating models in the
determination of the evolutionary masses. However, they find that the effect of
rotation should be negligible for objects with high gravities (and then,
early evolutionary stage). This is probably the case of HD~93205,
which is a member of a very young open cluster and does not exhibit
traces of evolution on its spectrum (for example, He{\sc ii}\,4686 \AA\
is seen as a deep absorption, like in stars near the Zero Age
Main Sequence). Moreover, the projected rotational velocities of
HD~93205, as found by Howarth et al (1997) are 135 km\,s$^{-1}$ and
145 km\,s$^{-1}$ for primary and secondary, respectively. With an
inclination of 54$^\circ$ -- 57$^\circ$ the actual rotational velocity of
the primary component would be 161 -- 167 km\,s$^{-1}$, which is not a
very large value.

The Carina Nebula, where HD~93205 is immersed,  harbours  many massive 
binaries. Among them we mention WR\,22, a remarkable Wolf-Rayet binary system 
containing the most massive WR star known at present in a binary.
 Recently, Schweickhardt et al. (1999) derived masses of $55\pm7$\,M$_\odot$
 and $21\pm2$\,M$_\odot$ for the
WN7ha and O8-O9.5~III-V binary components of WR\,22. 
The similarity between these 
mass values and those estimated for HD~93205 is so striking that we wonder
 if both systems represent somewhat different evolutionary stages of the
 same kind of massive binary systems.
Even nearby $\eta$ Carin\ae\ may be a binary with components of 
 `only' $\sim 70 + 70 M_{\odot}$ (Damineli, Conti \& Lopes, 1997).
While the final jury is still
out, we suspect that well separated, unevolved binaries like HD 93205
provide the most reliable technique for determining stellar masses,
relying on a minimum of assumptions compared to most other techniques.
Hence, we are led to speculate that there are still serious problems of
overestimation of the masses of massive stars, based on both
evolutionary and spectroscopic methods based other than on Keplerian
orbits, that become progressively worse for the higher masses.

We have also found that HD~93205 presents fast apsidal rotation with
 a period of 
$185\pm16$ years mainly due to tidal forces. In a forthcoming paper 
(Benvenuto et al. 2000) we will try to use this new determination to find 
an independent estimate of the masses of the binary components of HD~93205.

While the total X-ray emission of HD~93205 results in a ratio 
$L_{x} / L_{bol}$ simlar to that observed in other O-type stars,
the X-ray lightcurve shows a rise in X-ray flux near periastron passage
probably due to wind wind collision effects. This effect is present in
both PSPC and HRI lightcurves, and the observed ratio of the flux at
periastron to the flux at apastron is about 1.7.

\bigskip

\section {Acknowledgements}

NIM, VSN, MAC \& JFAC acknowledge use at CASLEO of the CCD and data acquisition
system supported under U.S. NSF grant AST-90-15827 to R.M. Rich and want to 
thank the director and staff of CASLEO for the use of their facilities
and kind hospitality during the observing runs.

   GR acknowledges financial support from the FNRS and from contract 
   P4/05 PAI (SSTC-Belgium) and the PRODEX XMM-OM Project.

TM, AFJM and NSL  thank 
 the Natural Sciences and Engineering Research Concil (NSERC) of
 Canada and the Fonds pour la Formation de Chercheurs et l'Aide
 \`a la Recherche (FCAR) of Qu\'ebec for financial support.

RHB acknowledges financial support from Fundaci\'on Antorchas
(Project No. 13783-5).

The authors want to thank an anonymous referee for helpful suggestions.

\end{document}

%% file: instr.tex
\begin{table*}
\centering
\begin{minipage}{150mm}
\caption{Instrumental configurations for different observing runs}
\label{instr}
\begin{tabular}{@{}lllllllll}
Id. & Date & Observat. & Tel. & Spectrogr. & CCD & $\lambda / \Delta(\lambda)$ & Spec. Range [\AA] &  No. Obs. \\
1  & Jan. 1995      & CASLEO & 2.15-m & REOSC     & Tek 1024    & 7500  & 3500--6000 &  4 \\
2  & Jan. 1995      & UTSO    & 0.6-m  & Garrison & PM 512 & 1000  & 4000--5000 & 9 \\ 
3  & Feb. 1997      & CASLEO & 2.15-m & REOSC     & Tek 1024    & 15000 & 3500--6000 & 15 \\ 
4  & March 1997     & ESO    & 1.5-m  & B\&C      & ESO CCD\#39 & 3000  & 3850--4800 &  6 \\
5  & June 1997	    & ESO    & 1.4-m  & CAT CES   & ESO CCD\#38 & 70000 & 4445--4495 &  5 \\
6  & Dec. 1997      & CASLEO & 2.15-m & B\&C      & PM 512      & 1000  & 3900--5000 &  2 \\
7  & Jan.-Feb. 1998 & CASLEO & 2.15-m & REOSC     & Tek 1024    & 15000 & 3500--6000 & 12 \\
8  & Feb. 1998      & CASLEO & 2.15-m & B\&C      & PM 512      & 1000  & 3900--5000 &  2 \\
9  & July 1998	    & ESO    & 1.4-m  & CAT CES   & ESO CCD\#38 & 70000 & 4461--4480 &  5 \\
10 & Jan.-Feb. 1999 & CASLEO & 2.15-m & REOSC     & Tek 1024    & 15000 & 3500--6000 &  8 \\
\end{tabular}
\end{minipage}
\end{table*}

%% file: vr_low.tex
\begin{table}
\caption{Low dispersion radial velocities of HD 93205}
\label{vr_low}
\begin{tabular}{@{}lcrrr}
HJD            & Phase  & \multicolumn{1}{c}{O3\,V Star}     & \multicolumn{1}{c}{O8\,V Star}     & Run \\
$2\,400\,000+$ & $\phi$ & \multicolumn{1}{c}{[km\,s$^{-1}$]} & \multicolumn{1}{c}{[km\,s$^{-1}$]} & \\ 
\\
49732.745      & 0.963 &  $122\pm\ \,6\,(2)$ & $-326\pm45\,(2)$ & 2 \\
49733.622      & 0.108 &  $-37\pm\ \,7\,(3)$ &  $100\pm21\,(3)$ & 2 \\
49734.728      & 0.289 & $-107\pm\ \,5\,(3)$ &  $202\pm12\,(3)$ & 2 \\
49736.697      & 0.613 &    $8\pm\ \,9\,(3)$ &   $33\pm11\,(3)$ & 2 \\
49737.706      & 0.779 &   $76\pm12\,(3)$    & $-150\pm39\,(3)$ & 2 \\
\\
49738.715      & 0.945 &  $191\pm\ \,7\,(3)$ & $-285\pm38\,(3)$ & 2 \\
49742.695      & 0.600 &    $6\pm22\,(3)$    &   $69\pm11\,(3)$ & 2 \\
49743.672      & 0.760 &   $60\pm\ \,7(3)$   & $-119\pm14\,(3)$ & 2 \\
49744.674      & 0.925 &  $221\pm10\,(3)$    & $-375\pm22\,(3)$ & 2 \\
50534.666      & 0.852 &  $145\pm\ \,2\,(3)$ & $-294\pm36\,(3)$ & 4 \\
\\
50535.661      & 0.015 &   $81\pm12\,(3)$    & $-191\pm31\,(3)$ & 4 \\
50536.656      & 0.179 & $-108\pm16\,(3)$    &  $219\pm19\,(3)$ & 4 \\
50537.687      & 0.349 &  $-93\pm\ \,8\,(3)$ &  $224\pm15\,(3)$ & 4 \\
50538.652      & 0.507 &  $-57\pm\ \,1\,(2)$ &   $38\ \ \ \ \ \ \,\,(1)$      & 4 \\
50539.653      & 0.672 &  $-16\pm\ \,4\,(2)$ &  $-55\ \ \ \ \ \ \,\,(1)$      & 4 \\
\\
50807.844      & 0.780 &  $-19\pm19\,(3)$    & $-211\ \ \ \ \ \ \,\,(1)$      & 6 \\
50811.872      & 0.443 &  $-92\pm27\,(3)$    &   $58\ \ \ \ \ \ \,\,(1)$      & 6 \\
50858.819      & 0.164 &  $-98\pm12\,(3)$    &  $148\pm15\,(2)$ & 8 \\
50861.831      & 0.659 &  $-19\pm13\,(3)$ & $-112\pm11\,(2)$ & 8 \\
\end{tabular}

\medskip
Notes: The meaning of the columns is as follows:
(1)  Heliocentric Julian Date; (2) orbital phase computed with the new
ephemeris (see section 4); (3) heliocentric radial velocity of the
primary component; (4) heliocentric radial velocity of the secondary
component; (5) observing run identification. - 
Quoted errors are standard deviations of the mean derived from the
average of several spectra obtained in succesive exposures 
(between parenthesis) in Run 2, and from the 
average of several absorption lines (between parenthesis) in each spectrum 
for Runs 4, 6 and 8.
\end{table}

%% file: vr_high.tex
\begin{table*}
\caption{High-resolution radial velocity measurements for HD~93205}
\label{vr_high}
\begin{tabular}{@{}lcrcrccrccrcr}
\hline
HJD & Phase & \multicolumn{5}{c}{He\,{\sc i} 4471} & \multicolumn{5}{c}{He\,{\sc ii} 4686} & Run    \\
$2\,400\,000+$ & $\phi$ & \multicolumn{5}{c}{[km\,s$^{-1}$]} & \multicolumn{5}{c}{[km\,s$^{-1}$]} & \\
  	   &      & Prim. & $(O-C)_1$ & Sec. & $(O-C)_1$ & $(O-C)_3$ &  ~~~~~~Prim. & $(O-C)_2$ & $(O-C)_3$ & Sec. & $(O-C)_2$  \\
\hline
 49738.853 & 0.968 & $+138$ & $-19$ & $-337$ & $+30$ & $+26$ & $+161$ & $+14$ & $+11$ & $-338$ &  $+1$ & 1 \\
 49739.855 & 0.133 & $-91$  &  $-6$ & $+166$ & $+11$ &  $+4$ &  $-64$ &  $+5$ &  $+9$ & $+174$ & $+11$ & 1 \\
 49741.849 & 0.461 & $-91$  & $-11$ & $+150$ &  $+5$ &   $0$ &  $-56$ &  $+8$ & $+12$ & $+147$ &  $-6$ & 1 \\
 49742.854 & 0.626 &        &       &  $+31$ & $+27$ & $+20$ &  $+15$ & $+16$ & $+24$ &        &       & 1 \\
 50494.863 & 0.305 &  $-100$ & $+19$ & $+233$ &  $+8$ &  $+5$ &  $-97$ &  $+3$ &  $+4$ & $+231$ &  $-6$ & 3 \\
\\
 50495.628 & 0.431 &  $-96$ &  $-7$ & $+169$ &  $+5$ &   $0$ &  $-70$ &  $+3$ &  $+6$ & $+184$ & $+11$ & 3 \\
 50495.746 & 0.451 &        &       & $+153$ &  $+2$ &  $-4$ &  $-55$ & $+12$ & $+16$ & $+137$ & $-23$ & 3 \\
 50495.864 & 0.470 & $-121$ & $-44$ & $+135$ &  $-3$ &  $-9$ &  $-49$ & $+12$ & $+16$ & $+194$ & $+48$ & 3 \\
 50496.715 & 0.610 &        &       &  $+33$ & $+13$ &  $+6$ &   $+5$ & $+13$ & $+21$ &        &       & 3 \\
 50497.751 & 0.780 &        &       & $-194$ &  $-5$ & $-11$ &  $+81$ &   $0$ &  $+8$ & $-193$ &  $-7$ & 3 \\
\\
 50497.864 & 0.799 & $+115$ & $+27$ & $-212$ &  $+5$ &   $0$ & $+103$ & $+10$ & $+18$ &        &       & 3 \\
 50498.599 & 0.920 & $+154$ & $-12$ & $-390$ &  $-4$ &  $-5$ & $+162$ &  $+4$ &  $+3$ & $-366$ &  $-3$ & 3 \\
 50498.722 & 0.940 & $+141$ & $-27$ & $-378$ & $+13$ & $+12$ & $+176$ & $+18$ & $+15$ & $-387$ & $-22$ & 3 \\
 50498.852 & 0.961 & $+163$ &  $+2$ & $-388$ & $-12$ & $-15$ & $+151$ &   $0$ &  $-3$ & $-340$ &  $+8$ & 3 \\
 50505.659 & 0.081 &        &       &  $+21$ &  $-8$ & $-20$ &  $-10$ &  $+4$ & $+12$ &  $+46$ &  $+9$ & 3 \\
\\
 50505.806 & 0.105 &        &       & $+108$ & $+10$ &  $+1$ &  $-38$ &  $+6$ & $+12$ &        &       & 3 \\
 50506.610 & 0.237 &        &       & $+249$ & $+17$ & $+13$ &  $-91$ & $+13$ & $+13$ & $+257$ & $+12$ & 3 \\
 50507.755 & 0.426 &  $-48$ & $+42$ & $+174$ &  $+6$ &  $+2$ &  $-80$ &  $-6$ &  $-3$ & $+179$ &  $+2$ & 3 \\
 50508.614 & 0.567 &        &       &        &       &       &   $-3$ & $+23$ & $+30$ &        &       & 3 \\
 50622.519 & 0.300 & $-163$ & $-46$ & $+222$ &  $-4$ &  $-7$ &        &       &       &        &       & 5 \\
\\
 50623.538 & 0.468 &  $-94$ & $-17$ & $+148$ &  $+8$ &  $+3$ &        &       &       &        &       & 5 \\
 50624.496 & 0.626 &        &       &   $-3$ &  $-7$ & $-15$ &        &       &       &        &       & 5 \\
 50625.530 & 0.796 & $+100$ & $+15$ & $-233$ & $-21$ & $-27$ &        &       &       &        &       & 5 \\
 50626.499 & 0.955 & $+199$ & $+34$ & $-404$ & $-21$ & $-23$ &        &       &       &        &       & 5 \\
 50841.720 & 0.351 & $-104$ & $+5$  & $+199$ &  $-8$ & $-12$ &  $-95$ &  $-3$ &  $-1$ & $+202$ & $-17$ & 7 \\
\\
 50842.737 & 0.519 &        &       &  $+85$ & $-16$ & $-23$ &  $-43$ &  $+1$ &  $+7$ &  $+95$ & $-12$ & 7 \\
 50843.804 & 0.694 &        &       &  $-54$ & $+18$ & $+11$ &   $+9$ & $-23$ & $-14$ &        &       & 7 \\
 50844.814 & 0.860 & $+108$ & $-24$ & $-343$ & $-29$ & $-32$ & $+115$ & $-17$ & $-12$ & $-300$ &  $+2$ & 7 \\
 50845.792 & 0.021 & $+110$ & $+28$ & $-180$ & $+25$ & $+13$ &  $+76$ &  $-6$ &  $-1$ & $-187$ &   $0$ & 7 \\
 50846.797 & 0.186 & $-107$ &  $+5$ & $+208$ &  $-7$ & $-11$ & $-104$ &  $-9$ &  $-7$ & $+217$ &  $-9$ & 7 \\
\\
 50847.745 & 0.342 & $-108$ &  $+3$ & $+182$ & $-29$ & $-33$ & $-100$ &  $-6$ &  $-4$ & $+236$ & $+13$ & 7 \\
 50848.806 & 0.517 &        &       &  $+89$ & $-14$ & $-20$ &  $-24$ & $+21$ & $+27$ &        &       & 7 \\
 50849.826 & 0.685 &        &       &  $-50$ & $+11$ &  $+4$ &  $+20$ &  $-7$ &  $+2$ &  $-64$ &  $-5$ & 7 \\
 50850.819 & 0.848 &  $+80$ & $-43$ & $-290$ &  $+4$ &  $+1$ & $+127$ &  $+3$ &  $+8$ & $-277$ &  $+8$ & 7 \\
 50851.754 & 0.002 &        &       & $-278$ &  $+1$ &  $-8$ & $+105$ &  $-7$ &  $-5$ & $-259$ &  $-3$ & 7 \\
\\
 50852.805 & 0.175 & $-109$ & $-1$  & $+205$ &  $-1$ &  $-6$ &  $-99$ &  $-7$ &  $-5$ & $+238$ & $+21$ & 7 \\
 50995.546 & 0.651 &        &       &  $-23$ &  $-1$ &  $-8$ &        &       &       &        &       & 9 \\
 50996.503 & 0.808 & $+103$ & $+9$  & $-232$ &  $-1$ &  $-6$ &        &       &       &        &       & 9 \\
 50998.510 & 0.138 & $-113$ & $-24$ & $+168$ &  $+4$ &  $-2$ &        &       &       &        &       & 9 \\
 50999.512 & 0.303 & $-125$ & $-8$  & $+228$ &  $+3$ &  $-1$ &        &       &       &        &       & 9 \\
\\
 51000.472 & 0.461 &  $-87$ & $-7$  & $+157$ & $+12$ &  $+7$ &        &       &       &        &       & 9 \\
 51208.772 & 0.719 &        &       & $-87$  & $+16$ &  $+9$ &  $+37$ &  $-8$ &  $+1$ & $-146$ & $-45$ & 10\\
 51209.697 & 0.871 & $+124$ & $-16$ & $-325$ &  $+5$ &  $+2$ & $+132$ &  $-6$ &  $-2$ & $-310$ &  $+7$ & 10\\
 51210.850 & 0.061 &        &       &  $-47$ &  $-4$ & $-17$ &   $+9$ &  $-7$ &  $+1$ &        &       & 10\\
 51211.825 & 0.221 & $-124$ & $-5$  & $+240$ & $+10$ &  $+7$ & $-107$ &  $-5$ &  $-4$ & $+225$ & $-17$ & 10\\
\\
 51215.867 & 0.886 & $+158$ & $+9$  & $-346$ &  $+4$ &  $+3$ & $+154$ &  $+9$ & $+11$ & $-332$ &  $+3$ & 10\\
 51216.746 & 0.030 &  $+90$ & $+26$ & $-171$ &  $-4$ & $-17$ &  $+77$ & $+10$ & $+16$ & $-152$ &   $0$ & 10\\
 51217.870 & 0.215 &  $-99$ & $+19$ & $+229$ &  $+1$ &  $-3$ &  $-93$ &  $+9$ & $+10$ & $+236$ &  $-4$ & 10\\	 
 51218.822 & 0.372 &        &       & $+195$ &  $-3$ &  $-7$ &  $-77$ & $+11$ & $+13$ & $+208$ &  $-1$ & 10\\
\hline
\end{tabular}
Note: $(O-C)_1$, $(O-C)_2$ and $(O-C)_3$ refer to the solutions for He{\sc i}, He{\sc ii} and `Combined',
presented in Table 4 and Figures 3 and 4.
\end{table*}

%% file: orb_sol.tex
\begin{table*}
\begin{minipage}{170mm}
\caption{Orbital elements of HD 93205 binary components derived from
radial velocities obtained in high-resolution spectrograms}
\label{orb_sol}
\begin{tabular}{@{}llllll}
                        & He\,{\sc i}\ 4471\AA\ & He\,{\sc ii}\ 4686\AA\ & Combined~~~~~~~~~~~~~~~~~ & CW76 & SL93 \\
\\
$a_1\ \sin i$\ [km]	& $(1.12 \pm 0.02)\,10^7$ & $(1.03 \pm 0.02)\,10^7$ & $(1.03 \pm 0.02)\,10^7$ & $(1.015\pm0.047)\,10^7$ & $(1.071\pm0.021)\,10^7$\\
$a_2\ \sin i$\ [km]	& $(2.43 \pm 0.03)\,10^7$ & $(2.41 \pm 0.02)\,10^7$ & $(2.44 \pm 0.02)\,10^7$ &  & $(2.457\pm0.077)\,10^7$ \\
\\
$K_1$\ [km s$^{-1}$]	& $144.5 \pm 2.6$ & $131.5 \pm 2.3$ & $132.6\pm2.0$ & $139.1\pm6.0$ & $141.9\pm3.0$ \\
$K_2$\ [km s$^{-1}$]	& $312.1 \pm 2.5$ & $308.3 \pm 2.3$ & $313.6\pm1.8$ & $360\pm53$ & $324.8\pm10.3$ \\
\\
$P$\ [days]		& $6.0803\pm 0.0004$ & $6.0803\pm 0.0004$ & $6.0803\pm0.0004$ & $6.08102\pm0.00066$ & $6.08205\pm0.00033$ \\
$e$			& $0.37  \pm 0.01$   & $0.35 \pm 0.01$ & $0.370\pm0.005$ & $0.49\pm0.03$ & $0.436\pm0.016$ \\
$\omega$\ [degrees]	& $52.0  \pm 1.3$    & $55.7 \pm 1.4$ &  $50.8\pm0.9$ & $12\pm3$ & $16.3\pm1.8$ \\
$T_0$\ [+2,400,000 HJD] & $50499.112 \pm 0.017$ & $50499.138\pm0.017$ & $50499.089\pm0.012$ & $42532.784\pm0.060$ & $44113.818\pm0.02$ \\
$T_{\rm Vmax}$		& $50498.702 \pm 0.017$ & $50498.679\pm0.017$ & $50498.695\pm0.012$\\
$\gamma$\ [km s$^{-1}$]	& $-8.8   \pm 1.3$       & $2.2\pm1.1$ & $-2.9\pm0.9$ & $3.6\pm2.5$ & $30.2\pm1.8$ \\
\\
$M_1\,\sin^3 i$ [M$_\odot$] & $33.0 \pm 1.6$    & $30.8\pm1.3$ & $31.5\pm1.1$ & $39$ & $32.6\pm2.6$ \\
$M_2\,\sin^3 i$ [M$_\odot$] & $15.3 \pm 1.5$    & $13.1\pm1.1$ & $13.3\pm1.1$ & $15$ & $14.2\pm0.9$ \\
$q(M_2/M_1)$		& $0.463 \pm 0.012 $    & $0.430\pm0.010$ & $0.423\pm0.009$ & $0.385\pm0.090$ & $0.437\pm0.009$ \\ 
$P.E.$\ [km s$^{-1}$] & 11.4 & 8.6 & 8.4 & & \\ 
\end{tabular}

\medskip
Notes: $T_{\rm Vmax}$ means the time of maximum radial velocity of the
primary component. $P.E.$ represents the probable error of the fit.
\end{minipage}
\end{table*}

%% file: omega.tex
\begin{table}
\caption{Longitude of periastrom derived from different radial velocities datasets}
\label{omega.tab}
\begin{tabular}{@{}lcrl}
Dataset       & $T$           & $\pm\Delta T$ & \multicolumn{1}{c}{$\omega$}     \\
              & [HJD]         &               & \multicolumn{1}{c}{[degrees]}    \\
\\ 
Conti \& Walborn (1976)   & $2\,442\,321$      & $214$ & $\ \, 5.7\pm6.1$   \\
Stickland \& Lloyd (1993) & $2\,444\,319$      & $511$ & $15.8\pm4.7$  \\
Levato et al. (1991)      & $2\,445\,783$      & $  5$ & $20.3\pm14.8$ \\
Run 2                     & $2\,449\,740$      & $  6$ & $37.0\pm12.0$ \\
Runs 3, 5, 9, 10          & $2\,450\,847$      & $322$ & $50.6\pm0.9$  \\  
\end{tabular}

\medskip
Note: $\Delta T$ stands for the duration, counted from $T$, from the
beginning and to the end, of each data-set.
\end{table}

%% file: ew.tex
\begin{table*}
\begin{minipage}{110mm}
\caption{Equivalent width measurements}
\label{ew}
\begin{tabular}{@{}llllll}
Star	   & ST & He\,{\sc i}\,4471 & He\,{\sc ii}\,4542 & He\,{\sc ii}\,4686 & References\\
HD\,73882  & O8\,V       & 0.62              & 0.45               & 0.55         & 1 \\
HD\,93250  & O3\,V((f))  & 0.07              & 0.68               & 0.54         & 1 \\
	   &             & $0.12\pm0.03$     & $0.68\pm0.04$      & $0.58\pm0.03$& 2 \\
Tr16-22    & O8\,V       & $0.65\pm0.05$     & $0.49\pm0.03$      & $0.54\pm0.03$& 2 \\
HD\,93205 Prim. & O3\,V  & $0.07\pm0.02$     & $0.63\pm0.04$      & $0.54\pm0.03$& 2 \\
\hspace{1.25cm} Sec.  & O8\,V  & $0.18\pm0.03$     & $0.11\pm0.03$      & $0.11\pm0.03$& 2 \\
\end{tabular}
\medskip

1. Mathys (1988)\\
2. This paper
\end{minipage}
\end{table*}

%% file: rosat.tex
\begin{table}
\caption{ROSAT observations of HD~93205 considered in the present analysis}
\label{rosat}
\begin{tabular}{llr}
   \\
Sequence & Observer & Exposure Time (s) \\
\\
     rh150037n00 &  J. PULS&        3352 \\
    rh202331n00 & The XMEGA Group & 47096 \\
   rh900385a02 &  J. SCHMITT &        523 \\
  rh900385a03 &  J. SCHMITT &        40556 \\
 rh900385n00 &  J. SCHMITT &        11528 \\
    rh900644n00 &  M. CORCORAN &        1721 \\
   rp200108n00 &  W. WALDRON &        1610 \\
  rp201262n00 &  A. PAULDRACH &        5665 \\
 rp900176a01 &  J. SWANK &        14544 \\
 rp900176n00 &  J. SWANK&        24321 \\
\end{tabular}
\end{table}